\documentclass{article}
\usepackage{iclr2027_conference,times}

\usepackage{amsmath,amssymb,bm}
\usepackage{booktabs}
\usepackage{graphicx}
\usepackage{float}
\usepackage{array}
\usepackage{tabularx}
\usepackage{xcolor}
\usepackage{tikz}
\usetikzlibrary{arrows.meta,positioning,fit,calc,shapes.geometric,shapes.symbols}
\usepackage{hyperref}
\usepackage{url}
\hypersetup{hidelinks}

\definecolor{entityblue}{HTML}{DCEBFA}
\definecolor{indexgreen}{HTML}{DFF2E1}
\definecolor{attentionorange}{HTML}{FCE8D5}
\definecolor{graphpurple}{HTML}{ECE2F6}
\definecolor{edgegray}{HTML}{555555}

\iclrfinalcopy

\newcommand{\methodtitle}{The Thousand-Graph Hypothesis}
\newcommand{\emptyedges}{E_{\mathrm{input}}^{\mathrm{rel}}=\varnothing}
\newcommand{\taskgraph}{G_q}
\title{The Thousand-Graph Hypothesis: A Testable Hypothesis of Task-Conditioned Relation Materialization in Repository-Level Code Reasoning}
\author{Fei Ding\\Alibaba Group\thanks{Corresponding author: \href{mailto:dignfei@gmail.com}{dignfei@gmail.com}}}

\begin{document}

\maketitle

\begin{abstract}
Large software repositories far exceed the context size that language models can effectively process in one shot. Training repository knowledge into a model is expensive, quickly stale, and difficult to keep current; on-demand retrieval is flexible but can still miss dispersed dependencies; explicit external code graphs add recurring costs for relation extraction, graph synchronization, and consistency maintenance. We propose \methodtitle and the \emph{Implicit Relation Materialization Hypothesis}: repository systems may persist only entities, while relevant entities entering context allow self-attention to organize task-specific connections transiently. We argue the same repository can induce different latent task graphs across tasks. To fit entity sets within bounded context, we use a two-layer repository entity index separating global positioning from local entity detail. In a controlled end-to-end SWE-bench Verified setting with DeepSeek-V4-Flash, the base system, one-layer index, and two-layer index obtain 92.1\%, 94.2\%, and 95.6\% success rates, respectively. The results indicate the two-layer setting improves success and that the full system can complete repository-level repair without pre-built entity relation edges. This is compatible with the hypothesis, but does not directly verify the latent graph realized inside the model. The reported deployment record (over half a year, more than 200 repositories, and several GitHub fixes) together with the currently auditable subset forms a practical long-term trace.
\end{abstract}

\section{Introduction}

Modern software repositories include many files, functions, interfaces, configurations, tests, data structures, and engineering constraints, often far beyond what a model can process in one context window. In addition, repositories are dynamic: commits, refactoring, migration, and requirement changes continuously reshape repository state. Repository-level code reasoning is therefore not only a one-step long-context problem, but also a question of how up-to-date repository knowledge reaches current reasoning.

Even with larger context windows, positional decay and evidence interference still reduce effective use of key signals~\citep{liu-etal-2024-lost}.

One route is to train repository knowledge into a model, but this is costly and becomes stale as code evolves. A second route retrieves local files on demand; this is flexible but can miss task-critical entities across files under limited steps. A third route is explicit repository graphs that externalize call, inheritance, usage, and dependency relations~\citep{liu-etal-2025-codexgraph,ICLR2025_4a4a3c19,chen-etal-2025-locagent}. Explicit graphs are precise and queryable, but repository changes shift maintenance to graph updates and consistency work.

RepoCoder, Repoformer, and Agentless improve accessibility with iterative retrieval, selective retrieval, and structural localization~\citep{zhang-etal-2023-repocoder,pmlr-v235-wu24a,Xia_2025}. They primarily improve \emph{where} to read; the remaining question is whether task-relevant entities are enough once they are in context, or whether relations still need to be pre-built.

Maintaining explicit graphs also incurs lifecycle cost. RPG-Encoder provides incremental graph updates; in one cross-commit analysis it reports full rebuild of 14.7M tokens and 633k incremental tokens~\citep{luo2026closingloopuniversalrepository}. RIG and Codebase-Memory similarly address edge invalidation, re-linking, and consistency checks after repository changes~\citep{chernyshahar2026repositoryintelligencegraphdeterministic,vogel2026codebasememorytreesitterbasedknowledgegraphs}.

We propose a falsifiable counterpoint: repository systems may only need to persist \emph{entities}, not pre-built entity edges. Once relevant entities are in context, self-attention can temporarily materialize task-specific relations. Different tasks on the same repository can therefore induce different latent task graphs, which we call \methodtitle.

This hypothesis raises a scaling challenge: even entity-only indexing can exceed context if repositories are large. We address this by a two-layer entity index that separates global routing from local entity focus. The interface supports persistent entities and zero pre-built edges, while still assembling task-specific inputs.

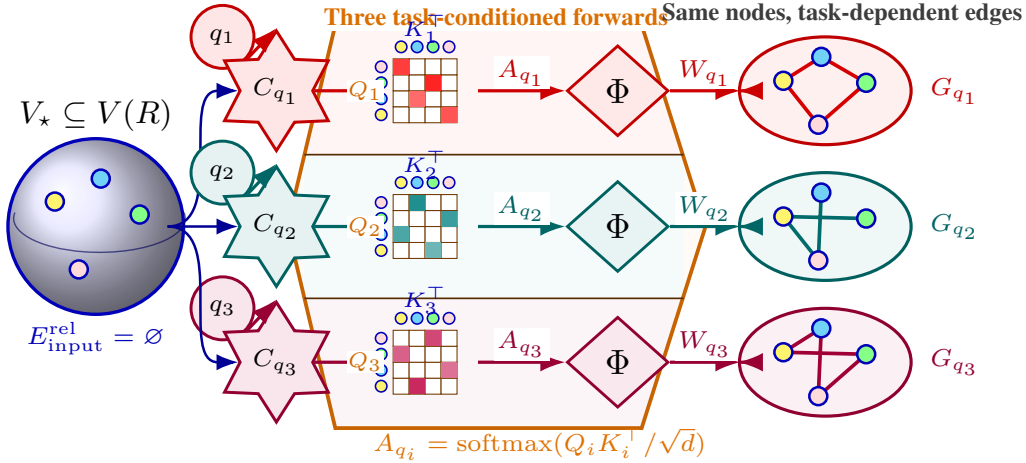
\begin{figure}[t]
\centering
\resizebox{0.99\linewidth}{!}{
\begin{tikzpicture}[
  >=Latex,
  entity/.style={circle,draw=blue!72!black,fill=white,
                 line width=0.72pt,minimum size=0.18cm,inner sep=0pt},
  task/.style={circle,draw=#1!76!black,fill=#1!10,
               line width=0.85pt,minimum size=0.54cm,font=\scriptsize},
  context/.style={star,star points=6,star point ratio=1.65,
                  draw=#1!76!black,fill=#1!10,line width=0.85pt,
                  minimum size=0.62cm,inner sep=0.6pt,font=\scriptsize},
  graph/.style={ellipse,draw=#1!75!black,fill=#1!6,
                minimum width=1.72cm,minimum height=1.08cm,line width=0.9pt},
  ray/.style={->,line width=0.92pt,draw=#1!80!black},
  eray/.style={->,line width=0.72pt,draw=blue!58!black},
  relation/.style={line width=1.05pt,draw=#1!84!black},
  note/.style={font=\scriptsize,fill=white,inner sep=1pt,align=center}
]

\shade[ball color=blue!11,opacity=0.96] (-4.75,0) circle (0.88cm);
\draw[blue!72!black,line width=0.95pt] (-4.75,0) circle (0.88cm);
\draw[blue!35!black,opacity=0.62] (-5.56,0.14)
      arc[start angle=170,end angle=370,x radius=0.86cm,y radius=0.34cm];
\node[font=\small] at (-4.75,1.12) {$V_\star\subseteq V(R)$};
\node[font=\scriptsize,text=blue!72!black] at (-4.75,-1.12)
      {$E_{\mathrm{input}}^{\mathrm{rel}}=\varnothing$};
\node[entity,fill=yellow!70] at (-5.16,0.25) {};
\node[entity,fill=cyan!48]   at (-4.70,0.48) {};
\node[entity,fill=green!48]  at (-4.31,0.12) {};
\node[entity,fill=pink!58]   at (-4.92,-0.43) {};

\coordinate (h1) at (-2.35,2.02);
\coordinate (h2) at (0.72,2.02);
\coordinate (h3) at (1.36,0);
\coordinate (h4) at (0.72,-2.02);
\coordinate (h5) at (-2.35,-2.02);
\coordinate (h6) at (-2.83,0);
\path[fill=orange!4] (h1)--(h2)--(h3)--(h4)--(h5)--(h6)--cycle;
\path[fill=red!4]    (h1)--(h2)--(1.14,0.72)--(-2.66,0.72)--cycle;
\path[fill=teal!4]   (-2.66,0.72)--(1.14,0.72)--(1.14,-0.72)--(-2.66,-0.72)--cycle;
\path[fill=purple!4] (-2.66,-0.72)--(1.14,-0.72)--(h4)--(h5)--cycle;
\draw[orange!78!black,line width=1.08pt]
      (h1)--(h2)--(h3)--(h4)--(h5)--(h6)--cycle;
\draw[orange!34!black,line width=0.48pt] (-2.66,0.72)--(1.14,0.72);
\draw[orange!34!black,line width=0.48pt] (-2.66,-0.72)--(1.14,-0.72);
\node[font=\scriptsize\bfseries,text=orange!86!black,fill=white,inner sep=1.6pt]
      at (-0.74,2.10) {Three task-conditioned forwards};
\node[font=\scriptsize\bfseries,text=orange!88!black]
      at (-0.30,-2.18)
      {$A_{q_i}=\operatorname{softmax}(Q_iK_i^\top/\sqrt d)$};

\foreach \i/\yy/\col in {1/1.36/red,2/0/teal,3/-1.36/purple}{
  \node[task=\col] (q\i) at (-3.48,\yy+0.54) {$q_\i$};
  \node[context=\col] (c\i) at (-2.94,\yy) {$C_{q_\i}$};
  \draw[ray=\col] (q\i.south east) -- (c\i.north);
  \draw[eray] (-4.03,0) to[out=0,in=180] (c\i.west);
  \draw[ray=\col] (c\i.east) -- (-1.94,\yy);
}
\newcommand{\attentionmatrix}[5]{%
  \begin{scope}[shift={(#1,#2)}]
    \foreach \r in {0,1,2,3}{
      \foreach \c in {0,1,2,3}{
        \path[draw=orange!45!black,fill=white,line width=0.30pt]
          (\c*0.16,\r*0.16) rectangle ++(0.16,0.16);
      }
    }
    \ifcase#5\relax
      \path[fill=#3!76] (0.00,0.48) rectangle ++(0.16,0.16);
      \path[fill=#3!58] (0.16,0.16) rectangle ++(0.16,0.16);
      \path[fill=#3!86] (0.32,0.32) rectangle ++(0.16,0.16);
      \path[fill=#3!67] (0.48,0.00) rectangle ++(0.16,0.16);
    \or
      \path[fill=#3!70] (0.00,0.16) rectangle ++(0.16,0.16);
      \path[fill=#3!88] (0.16,0.48) rectangle ++(0.16,0.16);
      \path[fill=#3!57] (0.32,0.00) rectangle ++(0.16,0.16);
      \path[fill=#3!77] (0.48,0.32) rectangle ++(0.16,0.16);
    \else
      \path[fill=#3!62] (0.00,0.32) rectangle ++(0.16,0.16);
      \path[fill=#3!84] (0.16,0.00) rectangle ++(0.16,0.16);
      \path[fill=#3!72] (0.32,0.48) rectangle ++(0.16,0.16);
      \path[fill=#3!55] (0.48,0.16) rectangle ++(0.16,0.16);
    \fi
    \foreach \p/\cc in {0/yellow!70,1/cyan!48,2/green!48,3/pink!58}{
      \node[entity,fill=\cc,minimum size=0.115cm,line width=0.45pt]
        at (\p*0.16+0.08,0.76) {};
      \node[entity,fill=\cc,minimum size=0.115cm,line width=0.45pt]
        at (-0.12,\p*0.16+0.08) {};
    }
    \node[font=\tiny,text=blue!74!black] at (0.32,0.94) {$K_{#4}^{\top}$};
    \node[font=\tiny,text=orange!82!black,fill=white,inner sep=0.4pt]
      at (-0.28,0.32) {$Q_{#4}$};
  \end{scope}
}
\attentionmatrix{-1.77}{1.04}{red}{1}{0}
\attentionmatrix{-1.77}{-0.32}{teal}{2}{1}
\attentionmatrix{-1.77}{-1.68}{purple}{3}{2}

\foreach \i/\yy/\col in {1/1.36/red,2/0/teal,3/-1.36/purple}{
  \node[diamond,aspect=1.12,draw=\col!76!black,fill=\col!9,
        line width=0.82pt,minimum width=0.62cm,minimum height=0.55cm]
        (phi\i) at (0.48,\yy) {$\Phi$};
  \draw[ray=\col] (-0.92,\yy) --
        node[note,above,text=\col!76!black] {$A_{q_\i}$} (phi\i.west);
  \draw[ray=\col] (phi\i.east) --
        node[note,above,text=\col!76!black] {$W_{q_\i}$} (1.71,\yy);
}

\foreach \i/\yy/\col in {1/1.36/red,2/0/teal,3/-1.36/purple}{
  \node[graph=\col] (g\i) at (2.56,\yy) {};
  \node[font=\scriptsize,text=\col!78!black,anchor=west]
        at (3.49,\yy) {$G_{q_\i}$};
  \draw[ray=\col] (1.71,\yy) -- (g\i.west);
  \coordinate (a\i) at (2.14,\yy+0.10);
  \coordinate (b\i) at (2.52,\yy+0.34);
  \coordinate (c\i) at (2.97,\yy+0.08);
  \coordinate (d\i) at (2.49,\yy-0.34);
}
\draw[relation=red]    (a1)--(b1)--(c1) (a1)--(d1)--(c1);
\draw[relation=teal]   (a2)--(c2) (a2)--(d2)--(b2);
\draw[relation=purple] (a3)--(b3)--(d3) (a3)--(c3)--(d3);
\foreach \i in {1,2,3}{
  \node[entity,fill=yellow!70] at (a\i) {};
  \node[entity,fill=cyan!48]   at (b\i) {};
  \node[entity,fill=green!48]  at (c\i) {};
  \node[entity,fill=pink!58]   at (d\i) {};
}
\node[font=\scriptsize\bfseries,text=black!76] at (2.73,2.13)
      {Same nodes, task-dependent edges};
\end{tikzpicture}
}
\caption{Task-conditioned attention interpretation of \methodtitle. The shared entity base and different task contexts induce different relation structures; $\Phi$ aggregates cross-token, layer, and head interactions for each task.}
\label{fig:hypothesis}
\end{figure}

Our contributions are:
\begin{itemize}
    \item A new task framing of repository knowledge as a long-tail, evolving-state problem.
    \item A new mechanism hypothesis: task-conditioned relation materialization without persistent external relations.
    \item A scalable two-layer repository entity interface that separates global and local entity access.
    \item Controlled end-to-end evidence on one model and one public benchmark.
    \item A practical long-term deployment note separating full reported usage from the currently auditable subset.
\end{itemize}

\section{The Continuous Knowledge Gap}

\subsection{Problem formulation}

Let the repository at time $t$ be $R_t$ and its persistent entity set be $V(R_t)$. Repositories evolve, so typically $R_t\neq R_{t+1}$. A repository task $q$ typically touches a subset of entities, but those entities can be scattered across many files and modules. Let the effective context budget be $B$; when the flat representation of entity candidates exceeds $B$, feeding all information at once becomes infeasible.

Importing $R$ into model parameters requires repeated updates and has high cost. Without training, local retrieval must rediscover needed entities of $q$, which may miss cross-file constraints. Explicit graphs shift the burden to relation extraction, graph updates, and consistency checks. The resulting tradeoff has three tracks:
\begin{enumerate}
    \item \textbf{Train into the model}: expensive and stale.
    \item \textbf{On-demand retrieval}: efficient but may miss distant evidence.
    \item \textbf{External explicit graph}: expensive long-term maintenance.
\end{enumerate}

We study:
\begin{quote}
\textbf{How can repository-level tasks be solved without repeated full training and without long-term external relation maintenance while still enabling task-specific organization of entities during inference?}
\end{quote}

\subsection{Persist entities, materialize relations transiently}

The external system extracts a repository entity set
\[
V(R)=\{v_1,v_2,\ldots,v_n\}.
\]
Each entity $v_i=(\operatorname{id}_i,x_i,m_i)$ includes content or summary $x_i$ and metadata $m_i$ (type, path, span, signature, responsibility tag). \methodtitle assumes that external state stores only entities, while relation state for inference is transient:
\[
\underbrace{V(R_t)}_{\text{persistent repository entities}}
\quad\longrightarrow\quad
\underbrace{G_q}_{\text{task graph formed in inference}},
\]
with $E_{\mathrm{input}}^{\mathrm{rel}}=\emptyset$ by design.

\section{Related Work}

\paragraph{Long-context and retrieval-based software agents.}
SWE-bench evaluates repository repair on real GitHub tasks and executable tests~\citep{ICLR2024_edac78c3}. Long-context windows increase reachable code but do not remove positional or noise-related degradation~\citep{liu-etal-2024-lost}. RepoCoder, Repoformer, and Agentless improve search via iterative retrieval, selective retrieval, or hierarchical structure~\citep{zhang-etal-2023-repocoder,pmlr-v235-wu24a,Xia_2025}, while still requiring retrieval orchestration. RLCoder retrieves completion-oriented context without persistent relation graphs, though training data still relies on dependency abstractions~\citep{Wang_2025}. This body addresses \emph{accessing the right files}; our question is whether relations must also be pre-computed.

\paragraph{Explicit repository graphs.}
GraphCoder, CodexGraph, RepoGraph, and LocAgent build graph-based navigation or retrieval structures~\citep{Liu_2024,liu-etal-2025-codexgraph,ICLR2025_4a4a3c19,chen-etal-2025-locagent}. GRACE, RepoScope, and CoCo maintain multi-view graphs and retrieve structured context, and RepoScope adds task-conditioned chain prediction~\citep{wang2025gracegraphguidedrepositoryawarecode,liu2025reposcopeleveragingchainawaremultiview,zhao2025completioncomprehensionguidingcode}. These methods are strong baselines for explicit-graph systems, but they still rely on external graph objects and maintenance.

\paragraph{Cost and lifecycle of explicit relations.}
Explicit graphs provide deterministic querying but require schema design, relation extraction, cross-file linking, storage, and update workflows. CodexGraph requires full scans and graph persistence~\citep{liu-etal-2025-codexgraph}; RPG-Encoder introduces incremental graph edits~\citep{luo2026rpgrepositoryplanninggraph}; RIG and Codebase-Memory address synchronization and stale-edge repair~\citep{chernyshahar2026repositoryintelligencegraphdeterministic,vogel2026codebasememorytreesitterbasedknowledgegraphs}.

\section{Two-Layer Repository Entity Interface}
\subsection{No persistent external relations}

The external state includes entities and indexes only. We define
\[
\emptyedges.
\]
Entity updates are applied to content and indexes; external relation edges are not persisted.

\subsection{Layer 1: global index}
Layer 1 returns a compact candidate set from repository-level modules and responsibilities:
\[
\mathcal{D}_q=\operatorname{Locate}(q,I^{(1)}(R)).
\]

\subsection{Layer 2: entity index}
Layer 2 selects local task-relevant entities in each candidate domain:
\[
V_q=\operatorname{Select}(q,\mathcal{D}_q,I^{(2)}(R))\subseteq V(R).
\]
The final task input is
\[
C_q=\operatorname{Serialize}(q,V_q),
\]
which contains only tasks and entity content, not pre-built relation edges.

\subsection{Workflow}

\begin{figure}[t]
\centering
\resizebox{0.98\linewidth}{!}{
\begin{tikzpicture}[
  >=Latex,
  beam/.style={line width=1.0pt,draw=blue!48!black},
  flow/.style={->,line width=1.08pt,draw=black!62},
  condition/.style={->,line width=0.90pt,dashed,draw=orange!78!black},
  feedback/.style={->,line width=1.00pt,dashed,draw=orange!84!black},
  dot/.style={circle,fill=blue!70!black,inner sep=1.25pt},
  lens/.style={ellipse,draw=green!58!black,fill=indexgreen,
               line width=1.05pt,minimum width=0.70cm,
               minimum height=#1},
  label/.style={font=\scriptsize,align=center,text=black!74}
]
\path[draw=blue!48!black,fill=entityblue!32,rounded corners=1pt]
      (-6.00,-0.68) rectangle (-4.28,0.74);
\path[draw=blue!54!black,fill=entityblue!54,rounded corners=1pt]
      (-5.80,-0.54) rectangle (-4.08,0.88);
\path[draw=blue!66!black,fill=entityblue,rounded corners=1pt,line width=0.95pt]
      (-5.60,-0.40) rectangle (-3.88,1.02);
\draw[blue!56!black] (-5.31,0.68)--(-4.20,0.68);
\draw[blue!56!black] (-5.31,0.36)--(-4.42,0.36);
\draw[blue!56!black] (-5.31,0.04)--(-4.28,0.04);
\node[label] at (-4.78,-0.90) {$R_t$};

\node[lens=2.30cm] (lens1) at (-2.82,0.30) {};
\node[lens=1.62cm] (lens2) at (-1.18,0.30) {};
\node[font=\small\bfseries,text=green!40!black] at (-2.82,0.43) {$I^{(1)}$};
\node[font=\small\bfseries,text=green!40!black] at (-1.18,0.43) {$I^{(2)}$};
\node[label,text=green!34!black] at (-2.82,-0.99) {Global routing};
\node[label,text=green!34!black] at (-1.18,-0.72) {Entity focus};
\node[circle,draw=orange!80!black,fill=orange!12,line width=0.9pt,
      minimum size=0.62cm,font=\small] (taskq) at (-2.00,1.67) {$q$};
\draw[condition] (taskq.south west) -- (lens1.north);
\draw[condition] (taskq.south east) -- (lens2.north);

\draw[beam] (-3.88,0.91) -- (-2.82,1.45) -- (-1.18,1.11) -- (0.16,0.67);
\draw[beam] (-3.88,-0.27) -- (-2.82,-0.85) -- (-1.18,-0.51) -- (0.16,-0.07);
\foreach \x/\y in {
  -3.69/0.64,-3.55/0.39,-3.68/0.14,-3.53/-0.09,-3.66/-0.30,
  -2.29/0.58,-2.06/0.18,-2.25/-0.22,
  -0.61/0.34,-0.42/-0.01}{
  \node[dot] at (\x,\y) {};
}

\node[star,star points=8,star point ratio=1.75,
      draw=orange!78!black,fill=yellow!22,line width=0.95pt,
      minimum size=1.10cm] (focus) at (0.78,0.30) {};
\node[font=\small] at (0.78,0.30) {$C_q$};
\node[label] at (0.78,-0.66) {Zero-edge focus};

\node[starburst,starburst points=13,starburst point height=0.13cm,
      draw=orange!84!black,fill=attentionorange,line width=1.0pt,
      minimum width=1.72cm,minimum height=1.58cm] (model) at (2.92,0.30) {};
\node[font=\large] at (2.92,0.30) {$F_\theta$};
\draw[->,orange!78!black,line width=0.8pt]
      ([shift={(130:0.47)}]model.center)
      arc[start angle=130,end angle=-105,radius=0.47];
\node[label] at (2.92,-0.78) {Reasoning loop};

\path[draw=purple!72!black,fill=graphpurple,line width=1.0pt]
      (4.40,0.12) -- (5.49,0.12) -- (5.49,1.12) --
      (5.18,1.43) -- (4.40,1.43) -- cycle;
\draw[purple!66!black] (5.18,1.43)--(5.18,1.12)--(5.49,1.12);
\draw[purple!54!black] (4.60,1.02)--(5.22,1.02);
\draw[purple!54!black] (4.60,0.71)--(5.28,0.71);
\node[font=\small] (patch) at (4.94,0.43) {$\Delta R$};
\node[label,anchor=west] at (5.58,0.52) {Patch};
\node[diamond,aspect=1.48,draw=teal!72!black,fill=teal!10,
      line width=1.0pt,minimum width=1.16cm,minimum height=0.82cm]
      (test) at (4.94,-1.04) {\scriptsize Test};
\node[circle,draw=green!62!black,fill=green!13,line width=0.9pt,
      minimum size=0.66cm] (accept) at (6.07,-1.04) {$\checkmark$};
\draw[flow] (-3.88,0.30) -- (lens1.west);
\draw[flow] (lens1.east) -- (lens2.west);
\draw[flow] (lens2.east) -- (focus.west);
\draw[flow] (focus.east) -- (model.west);
\draw[flow] (model.east) -- (4.40,0.67);
\draw[flow] (patch.south) -- (test.north);
\draw[feedback] (test.west) to[out=195,in=-72]
      node[label,below,text=orange!86!black,pos=0.52]{Failure: new evidence}
      (model.south);
\draw[->,green!58!black,line width=0.95pt] (test.east) --
      node[label,above,text=green!45!black]{Pass} (accept.west);
\draw[->,blue!48!black,dashed,line width=0.76pt]
      (accept.east)
      .. controls (6.98,-0.96) and (6.98,2.27) .. (5.75,2.27)
      -- node[label,above,yshift=1pt,text=blue!52!black,pos=0.54]
      {Update entities and indices only}
      (-4.58,2.27)
      .. controls (-4.93,2.27) and (-4.87,1.45) .. (-4.78,1.02);
\end{tikzpicture}
}
\caption{Task telescope for two-layer entity interface. Task $q$ narrows repository entities through two lenses to context $C_q$ without persistent relation edges.}
\label{fig:method}
\end{figure}
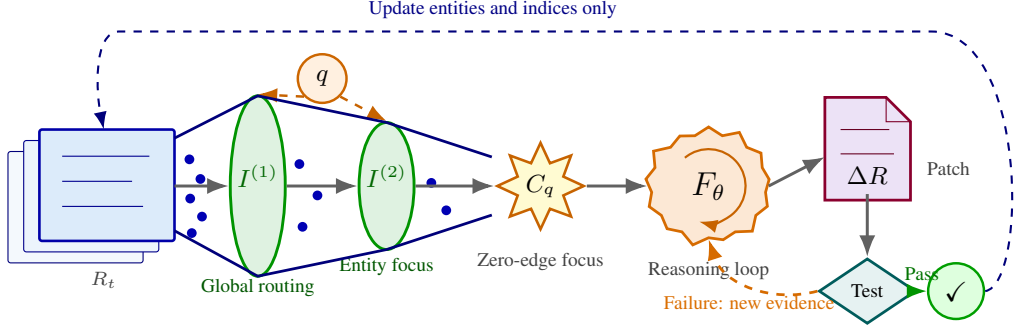

Figure~\ref{fig:method} summarizes the process: persistent extraction, global routing, local entity selection, prompt assembly, and reasoning. Patch validation runs through tests; only successful passes are written back to index updates, while entity and relation edges are not represented externally.

Compared with explicit-graph systems that maintain $(V,E)$, this work keeps only $V$ and indexes:
\begin{table}[t]
\caption{Boundary comparison between explicit external graphs and this approach}
\label{tab:comparison}
\centering
\small
\begin{tabularx}{\linewidth}{>{\raggedright\arraybackslash}p{1.15cm}
  >{\raggedright\arraybackslash}X
  >{\raggedright\arraybackslash}X}
\toprule
Dimension & Explicit graph & This work \\
\midrule
External state & Entities, edges, schema, graph database & Two-layer entity index \\
Where relations come from & External extraction or inference over source graph & Hypothesis: task-conditioned attention \\
Task adaptation & Fixed graph with local subgraph query & Task-specific transient organization in inference \\
Repository updates & Recompute entities and edges with consistency & Update entities and indexes only \\
\bottomrule
\end{tabularx}
\end{table}

\section{\methodtitle: Implicit Relation Materialization}
\subsection{From self-attention to task-conditioned relation}

Let $V_{q_1}=V_{q_2}=V_{q_3}=V_\star\subseteq V(R)$ in Fig.~\ref{fig:hypothesis} isolate task identity. In practice, tasks still use two-layer indexing for $V_q$; each panel is simplified.

For layer $\ell$ and head $h$:
\[
A_q^{(\ell,h)}
=
\operatorname{softmax}\left(
\frac{Q_q^{(\ell,h)}K_q^{(\ell,h)\top}}{\sqrt{d}}
\right).
\]
For entity token sets $T(v_i)$ and $T(v_j)$, define
\[
w_{ij}(q)=\Phi\left(\left\{A_q^{(\ell,h)}[a,b]\,\middle|\,a\in T(v_i),\,b\in T(v_j),\,\ell,h\right\}\right),
\]
and the task graph
\[
\taskgraph=(V_q,W_q),\quad W_q=\{w_{ij}(q)\}.
\]
$W_q$ is not an explicitly stored external graph; it is a hypothesis-defined latent quantity. We do not output or persist such edges.

\subsection{Hypothesis}

For different tasks $q_1$ and $q_2$ in the same repository, we predict:
\[
G_{q_1}\neq G_{q_2}.
\]
This is a behavioral prediction, not a deterministic identity from input differences alone.
\begin{quote}
\textbf{Repository-level systems can persist only entities while task-specific latent relations are materialized during inference by context conditions.}
\end{quote}

\subsection{Falsifiable predictions}
\begin{enumerate}
    \item \textbf{Task sensitivity.} Relevant entities and interactions differ by task.
    \item \textbf{Zero-edge feasibility.} Without pre-built edges, repository repair can still succeed if task entities enter context.
    \item \textbf{Coverage effect.} Missing entity coverage hurts success; higher coverage improves it.
\end{enumerate}
Our experiments directly test the second and indirectly the third through layer ablations.

\section{Controlled Experiment}
\subsection{Design}

We evaluate three conditions on SWE-bench Verified with DeepSeek-V4-Flash:
\begin{itemize}
    \item Base system: no entity index
    \item One-layer index
    \item Two-layer index
\end{itemize}
All conditions are controlled to have no pre-built entity-relation edges.

\begin{table}[t]
\caption{DeepSeek-V4-Flash on SWE-bench Verified.}
\label{tab:main_results}
\centering
\begin{tabular}{lcc}
\toprule
Condition & Pre-built relation edges & Success rate (\%) \\
\midrule
Base system & none & 92.1 \\
One-layer index & none & 94.2 \\
Two-layer index & none & \textbf{95.6} \\
\bottomrule
\end{tabular}
\end{table}

As shown in Table~\ref{tab:main_results}, the two-layer design improves success from 92.1\% to 95.6\%, with an absolute gain of 3.5 points over the base and 1.4 points over one-layer. This corresponds to an error reduction from 7.9\% to 4.4\% versus base and from 5.8\% to 4.4\% versus one-layer.

\section{Long-term Practice}

The authors report over half a year of usage, participation in more than 200 projects, and multiple public GitHub fixes. The currently auditable subset includes 6 repositories, 52 validation artifacts, an upper-bound public baseline date of 2026-07-29, and 29 merged PRs in one open repository (27 with explicit fix commits). It supports practical deployment viability but is not a direct proof of causal effect.

\begin{table}[H]
\caption{Reported scale vs. auditable subset}
\label{tab:practice}
\centering
\small
\begin{tabularx}{\linewidth}{>{\raggedright\arraybackslash}p{2.4cm}X}
\toprule
Evidence level & Records \\
\midrule
Author-reported overall & Half-year-plus usage; 200+ projects; several GitHub fixes \\
Auditable local snapshot & 6 repositories, 52 artifacts \\
Public GitHub subset & 1 repository; 29 merged PRs; 27 explicit-fix PRs \\
Index evolution in repair & 19 index-sync commits; 11 close index+fix sequences \\
\bottomrule
\end{tabularx}
\end{table}

\section{Conclusion and Boundary}

We define the repository knowledge gap and propose \methodtitle as a mechanism hypothesis: persistent entities, task-time relation materialization, and no prebuilt external relation graph. In this controlled setting, two-layer indexing attains 95.6\% success and outperforms one-layer and base conditions under zero-edge constraints. The current scope does not directly establish internal causal mechanisms, cross-model generality, or full long-term cost accounting.

\section*{AI Statement}
This paper used generative AI tools for drafting and refinement support, including wording polish, falsifiability checks, LaTeX style adjustments, and figure preparation. The core ideas, experimental methods, implementations, and results were provided and approved by the authors; AI tools did not replace experiments or alter primary observed outcomes.

\section*{Reproducibility Statement}
We report the base model, public benchmark, and three controlled conditions. The release package will include index-construction scripts, prompts, configurations, prediction logs, and SWE-bench run logs for the auditable subset.

\end{document}